\documentclass[preprint,aps,tightenlines,showpacs]{revtex4}

\usepackage{graphicx,epsfig}

\usepackage{dcolumn}     

\renewcommand{\case}{\frac}
\def\beq{\begin{equation}}
\def\eeq{\end{equation}}
\def\beqy{\begin{eqnarray}}
\def\eeqy{\end{eqnarray}}

\begin{document}

{

\title{Quantum Monte Carlo calculations of electroweak transition matrix
elements in $A = 6,7$ nuclei}

\author{Muslema Pervin}
\email{muslema@phy.anl.gov}
\author{Steven C. Pieper}
\email{spieper@anl.gov}
\author{R. B. Wiringa}
\email{wiringa@anl.gov}
\affiliation{Physics Division, Argonne National Laboratory,
Argonne, Illinois 60439}

\date{\today}

\begin{abstract}
Green's function Monte Carlo calculations of magnetic dipole, electric quadrupole,
Fermi, and Gamow-Teller transition
matrix elements are reported for $A=6,7$ nuclei.
The matrix elements are extrapolated from mixed estimates that bracket the
relevant electroweak operator between variational Monte Carlo and GFMC
propagated wave functions.
Because they are off-diagonal terms, two mixed estimates are required for
each transition, with a VMC initial (final) state paired with a GFMC final
(initial) state.
The realistic Argonne $v_{18}$ two-nucleon and Illinois-2 three-nucleon
interactions are used to generate the nuclear states.
In most cases we find good agreement with experimental data.
\end{abstract}


\maketitle

}

\section {Introduction}

The variational Monte Carlo (VMC) and Green's function Monte Carlo (GFMC)
techniques are powerful tools for calculating properties of light nuclei.
The GFMC method, in combination with the Argonne $v_{18}$ (AV18) two-nucleon
($N\!N$) and Illinois-2 (IL2) three-nucleon ($3N$) potentials, reproduces
the ground- and excited-state energies for $A\le10$ nuclei
\cite{PW01, PVW02, PWC04, P05}.
It is beginning to be used to calculate reactions, such as nucleon-nucleus
scattering~\cite{NPWCH07}.
Electroweak transitions in $A=6,7$ nuclei have been calculated using the
more approximate VMC technique with AV18 and the older Urbana-IX (UIX) 3N
potential.
These include $^6$Li elastic and transition form factors and radiative
widths~\cite{WS98}, electric quadrupole ($E2$) transition probabilities in $^7$Li for pion
inelastic scattering~\cite{LW01}, Gamow-Teller (GT) matrix elements in
$^6$He and $^7$Be weak decays~\cite{SW02}, and radiative capture reactions
producing $^6$Li, $^7$Li, and $^7$Be~\cite{NWS01,N01}.

The VMC results for electroweak transitions reported in earlier works were
in reasonable agreement with experimental data.
However, the AV18+IL2 Hamiltonian reproduces $p$-shell binding energies
better than AV18+UIX, and GFMC wave functions are better approximations to
the true eigenstates.
Hence GFMC calculations with AV18+IL2 for these electroweak transitions
are worth investigating.
In this work we study electromagnetic [$E2$ and
magnetic dipole ($M1$)] transition strengths and nuclear beta-decay
[Fermi (F) and GT] rates, with the GFMC technique using the
AV18+IL2 potential for nuclei with $A=6,7$.
This is the first off-diagonal matrix element calculation using the
nuclear GFMC method.
The GFMC technique of evaluating off-diagonal matrix elements should be
applicable in many other nuclear calculations, such as isospin-mixing,
and low-energy nuclear reactions.

In this work we consider only the
one-body parts of the transition operators.
Schematic expressions for these are
\beqy
E2&=& e{\sum_{k} } \frac{1}{2}\left[r_k^2 Y_2(\hat{r}_k)\right](1+\tau_{kz}),\\
M1&=& \mu_N {\sum_{k}} [(L_k+ g_p S_k)(1+\tau_{kz})/2 +g_n S_k(1-\tau_{kz})/2],\\
{\rm F}&=&{\sum_{k}} \tau_{k\pm},\\
{\rm GT}&=&{\sum_{k}} {\bf \sigma}_{k}\tau_{k\pm},
\eeqy
where k labels individual nucleons, r is the spatial coordinate, Y is a spherical harmonic, $\tau_{kz(\pm)}$ is the third
(raising/lowering) component of the isospin operator, $\sigma$ is the Pauli spin matrix, $\mu_N$ is the nuclear
magneton, $L (S)$ is the orbital (spin) angular momentum operator, and
$g_{p(n)}$ is the gyromagnetic ratio for a proton (neutron).
In all cases the appropriate projection ($z$, $+$, and $-$) of the
operators is understood.
Note that $e$, $\mu_N$, and $g_{p(n)}$ are all physical values, i.e., no
effective charge is used.

The current work does not include the contributions of two-body electroweak
current operators.
The effect of two-body meson-exchange operators has been studied previously,
and found to be small in electric quadrupole and weak transitions, but
significant for magnetic transitions~\cite{WS98,SW02}.

\section {VMC trial functions}
\label{sec:VMCwaves}

In this work we calculate the off-diagonal transition matrix element
$\langle \Psi_f(J^{\pi\prime};T^\prime)|{\cal O}| \Psi_i(J^\pi;T)\rangle$,
where $\cal O$ is one of the one-body electroweak transition operators
given above and $\Psi(J^\pi;T)$ is the nuclear wave function with a
specific spin-parity $J^\pi$ and isospin $T$.
The variational wave function is an approximate solution of the
many-body Schr\"{o}dinger equation
\beqy
 H \Psi(J^\pi;T)= E \Psi(J^\pi;T) \ .
\eeqy
The Hamiltonian used here has the form
\beqy
 H = \sum_{i} K_i  +  {\sum_{i<j}} v_{ij} +  \sum_{i<j<k} V_{ijk} \ ,
\eeqy
with $K_i$ being the non-relativistic kinetic energy, Argonne~$v_{18}$ 
\cite{WSS95} is the $N\!N$ potential $v_{ij}$, and Illinois-2
\cite{PPWC01} is the $3N$ interaction $V_{ijk}$.

The VMC trial function $\Psi_T(J^\pi;T)$ for a given nucleus is constructed
from products of two- and three-body correlation operators acting on an
antisymmetric single-particle state with the appropriate quantum numbers.
The correlation operators are designed to reflect the influence of the
interactions at short distances, while appropriate boundary conditions
are imposed at long range~\cite{W91,PPCPW97,NWS01,N01}.
The $\Psi_T(J^\pi;T)$ has embedded variational parameters
that are adjusted to minimize the expectation value
\begin{equation}
 E_V = \frac{\langle \Psi_T | H | \Psi_T \rangle}
            {\langle \Psi_T   |   \Psi_T \rangle} \geq E_0 \ ,
\label{eq:expect}
\end{equation}
which is evaluated by Metropolis Monte Carlo integration~\cite{MR2T2}.

A good variational trial function has the form
\begin{equation}
   |\Psi_T\rangle = \left[1 + \sum_{i<j<k}\tilde{U}^{TNI}_{ijk} \right]
                    \left[ {\cal S}\prod_{i<j}(1+U_{ij}) \right]
                    |\Psi_J\rangle \ .
\label{eq:psit}
\end{equation}
The Jastrow wave function, $\Psi_J$, is fully antisymmetric and has the
$(J^\pi;T)$ quantum numbers of the state of interest.
For s-shell nuclei, $\Psi_J$ has the simple form
\begin{equation}
    |\Psi_J\rangle = \left[ \prod_{i<j<k\leq A}f^c_{ijk} \right]
                     \left[ \prod_{i<j\leq A}f_c(r_{ij}) \right]
                    |\Phi_A(JM_JTT_{z})\rangle \ .
\label{eq:jastrow}
\end{equation}
Here $f_c(r_{ij})$ and $f^c_{ijk}$ are central two- and three-body correlation
functions and $\Phi_A$ is a Slater determinant in spin-isospin space, e.g.,
for the $\alpha$-particle, $|\Phi_{4}(0 0 0 0) \rangle
= {\mathcal A} |p\uparrow p\downarrow n\uparrow n\downarrow \rangle$.
The $U_{ij}$ and $\tilde{U}^{TNI}_{ijk}$ are noncommuting two- and
three-nucleon correlation operators, and ${\cal S}$
indicates a symmetric product over all possible ordering of pairs and triples.
The $U_{ij}$ includes spin, isospin, and tensor terms:
\begin{equation}
   U_{ij} = \sum_{p=2,6} u_p(r_{ij}) O^p_{ij} \ ,
\end{equation}
where the $O^{p=1,6}_{ij}=[1, {\bf\sigma}_{i}\cdot{\bf\sigma}_{j}, S_{ij}]
\otimes[1,{\bf\tau}_{i}\cdot{\bf\tau}_{j}]$ are the main static operators
that appear in the $N\!N$ potential.
The $f_c(r)$ and $u_p(r)$ functions are generated by the solution of a
set of coupled differential equations containing the bare $N\!N$ potential
with asymptotically-confined boundary conditions~\cite{W91}.
The $\tilde{U}^{TNI}_{ijk}$ has the spin-isospin structure of the dominant
parts of the $3N$ interaction as suggested by perturbation theory.

For the $p$-shell nuclei, $\Psi_J$ includes a one-body part that consists of
4 nucleons in an $\alpha$-like core and $(A-4)$ nucleons in $p$-shell orbitals.
We use $LS$ coupling to obtain the desired $JM_J$ value,
as suggested by standard shell-model studies~\cite{CK65}.
We also need to sum over different spatial symmetries $[n]$ of the angular
momentum coupling of the $p$-shell nucleons~\cite{BM69}.
The one-body parts are multiplied by products of central pair
and triplet correlation functions, which depend upon the shells ($s$ or $p$)
occupied by the particles and on the $LS[n]$ coupling:
\begin{eqnarray}
 |\Psi_{J}\rangle &=& {\cal A} \left\{\right.
   \prod_{i<j<k} f^{c}_{ijk}
   \prod_{i<j \leq 4}f_{ss}(r_{ij})
   \prod_{k \leq 4<l \leq A} f_{sp}(r_{kl})  \nonumber\\
                 && \sum_{LS[n]}
   \Big( \beta_{LS[n]} \prod_{4<l<m \leq A} f^{LS[n]}_{pp}(r_{lm})
  |\Phi_{A}(LS[n]JM_JTT_{3})_{1234:5\ldots A}\rangle \Big) \left.\right\} \ .
\label{eq:jastrow-p}
\end{eqnarray}
The operator ${\cal A}$ indicates an antisymmetric sum over all possible
partitions into 4 $s$-shell and $(A-4)$ $p$-shell particles.

The $LS[n]$ components of the single-particle wave function are given by:
\begin{eqnarray}
 &&  |\Phi_{A}(LS[n]JM_JTT_{3})_{1234:5\ldots A}\rangle =
   |\Phi_{4}(0 0 0 0)_{1234} \prod_{4 < l\leq A}
   \phi^{LS[n]}_{p}(R_{\alpha l}) \nonumber \\
 &&  \left\{ [ \prod_{4 < l\leq A} Y_{1m_l}(\Omega_{\alpha l}) ]_{LM_L[n]}
   \times [ \prod_{4 < l\leq A} \chi_{l}(\case{1}{2}m_s) ]_{SM_S}
   \right\}_{JM_J}
   \times [ \prod_{4 < l\leq A} \nu_{l}(\case{1}{2}t_3) ]_{TT_3}\rangle
    \ .
\end{eqnarray}
The $\phi^{LS}_{p}(R_{\alpha l})$ are $p$-wave solutions of a particle
in an effective $\alpha$-$N$ potential that has Woods-Saxon and Coulomb parts.
They are functions of the distance between the center of mass
of the $\alpha$ core and nucleon $l$, and may vary with $LS[n]$.
The $f_{ss}$, $f_{sp}$, and $f^{LS[n]}_{pp}$ all have similar short-range
behavior, like the $f_c$ of the $\alpha$-particle, but different long-range
tails.

Two different types of $\Psi_J$ have been constructed in recent VMC
calculations of light $p$-shell nuclei: an original shell-model kind of trial
function~\cite{PPCPW97} which we will call Type~I, and a cluster-cluster
kind of trial function~\cite{NWS01,N01} which we will call Type~II.
In Type~I trial functions, the $\phi^{LS}_{p}(r)$ has an exponential decay
at long range, with the depth, range, and surface thickness of the
Woods-Saxon potential serving as variational parameters.
The $f_{sp}$ goes to a constant $\sim 1$, while $f^{LS[n]}_{pp}$ has a
much smaller tail to allow clusterization of the $p$-shell nucleons.
Details of these $A=6,7$ trial functions are given in Ref.~\cite{PPCPW97}.

In Type~II trial functions, $\phi^{LS}_{p}(r)$ is again the solution of a
$p$-wave differential equation with a potential containing Woods-Saxon and
Coulomb terms, but with an added Lagrange multiplier that turns on at
long range.
This Lagrange multiplier imposes the boundary condition
\begin{equation}
\label{eqn:asymptotic}
[\phi^{LS[n]}_{p}(r\rightarrow\infty)]^{(A-4)} \propto W_{km}(2\gamma r)/r,
\end{equation}
where the Whittaker function, $W_{km}(2\gamma r)$, gives the asymptotic form of the
bound-state wave function in a Coulomb potential.
The $\gamma$ is related to the cluster separation energy which is taken from
experiment (the GFMC computed separation energies for AV18+IL2 are close to the experimental values).
The accompanying $f_{sp}$ goes to unity (more rapidly than in the
Type~I trial function) and the $f^{LS[n]}_{pp}$ are taken from the exact
deuteron wave function in the case of $^6$Li, or the VMC triton ($^3$He)
trial function in the case of $^7$Li ($^7$Be).
Consequently, the Type~II trial function factorizes at large cluster
separations as
\begin{equation}
\label{eqn:type2}
\Psi_T \rightarrow \psi_{\alpha} \psi_\tau
                 W_{km}(2\gamma r_{\alpha\tau})/r_{\alpha\tau} \ .
\end{equation}
where $\psi_{\alpha}$ and $\psi_\tau$ are the wave functions of the clusters
and $r_{\alpha\tau}$ is the separation between them.
More details on these wave functions are given in Refs.~\cite{NWS01,N01}.
In the case of $^6$He, which does not have an asymptotic two-cluster
threshold, we generate a $f^{LS[n]}_{pp}$ correlation assuming a
weakly-bound $^1$S$_0$ $nn$ pair.

For either type of trial function, a diagonalization is carried out in the
one-body basis to find the optimal values of the $\beta_{LS[n]}$ mixing
parameters for a given $(J^\pi;T)$ state.
The trial function, Eq.(\ref{eq:psit}), has the advantage of being efficient
to evaluate while including the bulk of the correlation effects.

\section {GFMC wave functions}
\label{gfmcwf}

The GFMC method~\cite{C87,C88} projects out the exact lowest-energy state,
$\Psi_{0}$, for a given set of quantum numbers, using
\begin{equation}
\Psi_0 = \lim_{\tau \rightarrow \infty} \exp [ - ( H^{\prime} - E_0) \tau ] \Psi_T \ ,
\label{eq:gfmc-1}
\end{equation}
where $H^{\prime}$ is a possibly simplified version of the 
desired Hamiltonian $H$ and
$\Psi_{T}$ is an initial trial wave function.
If the maximum $\tau$ actually used is large enough,
the eigenvalue $E_{0}$ is calculated exactly while other expectation values
are generally calculated neglecting terms of order $|\Psi_{0}-\Psi_{T}|^{2}$
and higher~\cite{PPCPW97}.
In contrast, the error in the variational energy $E_{V}$ is of order
$|\Psi_{0}-\Psi_{T}|^{2}$, and other expectation values calculated with
$\Psi_{T}$ have errors of order $|\Psi_{0}-\Psi_{T}|$.
In the following we present a brief overview of nuclear GFMC methods;
much more detail may be found in Refs.~\cite{PPCPW97,WPCP00}.

We start with the $\Psi_{T}$ of Eq.(\ref{eq:psit}) and define the propagated
wave function $\Psi(\tau)$
\begin{eqnarray}
 \Psi(\tau) = e^{-(H^{\prime}-E_{0})\tau} \Psi_{T}
            = \left[e^{-(H^{\prime}-E_{0})\triangle\tau}\right]^{n} \Psi_{T} \ ,
\end{eqnarray}
where we have introduced a small time step, $\tau=n\triangle\tau$;
obviously $\Psi(\tau=0) =  \Psi_{T}$ and
$\Psi(\tau \rightarrow \infty) = \Psi_{0}$.
Quantities of interest are evaluated in terms of a ``mixed'' expectation value
between $\Psi_T$ and $\Psi(\tau)$:
\begin{eqnarray}
\langle O(\tau) \rangle_M & = & \frac{\langle \Psi(\tau) | O |\Psi_{T}
\rangle}{\langle \Psi(\tau) | \Psi_{T}\rangle}.
\label{eq:expectation}
\end{eqnarray}
The desired expectation values would, of course, have $\Psi(\tau)$ on both
sides; by writing $\Psi(\tau) = \Psi_{T} + \delta\Psi(\tau)$  and neglecting
terms of order $[\delta\Psi(\tau)]^2$, we obtain the approximate expression

\begin{eqnarray}
\langle O (\tau)\rangle =
\frac{\langle\Psi(\tau)| O |\Psi(\tau)\rangle}
{\langle\Psi(\tau)|\Psi(\tau)\rangle}
\approx \langle O (\tau)\rangle_M
    + [\langle O (\tau)\rangle_M - \langle O \rangle_V] ~,
\label{eq:pc_gfmc}
\end{eqnarray}
where $\langle O \rangle_{\rm V}$ is the variational expectation value.
More accurate evaluations of $\langle O (\tau)\rangle$ are possible~\cite{K67},
essentially by measuring the observable at the mid-point of the GFMC propagation.
However, such estimates require a propagation twice as long as the mixed
estimate and require separate propagations for every expectation value
to be evaluated.  The nuclear calculations published to date use the
approximation of Eq.(\ref{eq:pc_gfmc}).

For off-diagonal matrix elements relevant to this work the mixed estimate is
generalized to the following expression
\begin{eqnarray}
 \frac{\langle\Psi^f(\tau)| O |\Psi^i(\tau)\rangle}{\sqrt{\langle \Psi^f(\tau) | \Psi^f(\tau)\rangle}
\sqrt{\langle \Psi^i(\tau) |\Psi^i(\tau)\rangle}}
\approx 
  \langle O(\tau) \rangle_{M_i}
+ \langle O(\tau) \rangle_{M_f}-\langle O \rangle_V \ ,
\label{eq:extrap}
\end{eqnarray}
where
\begin{eqnarray}
\label{eq:off}
\langle O \rangle_V 
& = & \frac{\langle \Psi^f_T | O |\Psi^i_{T}\rangle}
     {\sqrt{\langle \Psi^f_T| \Psi^f_T\rangle}
      \sqrt{\langle \Psi^i_T| \Psi^i_{T}\rangle}} \ , \\
\langle O(\tau) \rangle_{M_i} 
& = & \frac{\langle \Psi^f_T | O |\Psi^i(\tau)\rangle}
           {\langle \Psi^i_T|\Psi^i(\tau)\rangle}
      \sqrt{\frac{\langle \Psi^i_T|\Psi^i_T\rangle}
           {\langle \Psi^f_T | \Psi^f_{T}\rangle}} \ , 
\label{eq:mixed_i}  \\
\langle O(\tau) \rangle_{M_f} 
& = & \frac{\langle \Psi^f(\tau) | O|\Psi^i_{T}\rangle} 
           {\langle\Psi^f(\tau)|\Psi^f_{T}\rangle}
      \sqrt{\frac{\langle \Psi^f_T|\Psi^f_T\rangle}
           {\langle \Psi^i_T | \Psi^i_{T}\rangle}}  \ ,
\end{eqnarray}
and the index $i$ ($f$) refers to the wave function of the
initial (final) nuclear state.
In our calculation the operator always acts on the trial wave function.
The first term of Eq.(\ref{eq:mixed_i}) is thus replaced by its conjugate
$\langle \Psi^i(\tau) | O^\dagger |\Psi^{f}_T\rangle$ in our real computation.
Note here the computation for each extrapolated transition requires two
independent GFMC propagations.

The quantities
$\langle \Psi^i(\tau)|O|\Psi^f_{T}\rangle /
 \langle\Psi^i(\tau)|\Psi^i_{T}\rangle$ and
$\langle \Psi^f(\tau)|O|\Psi^i_{T}\rangle /
\langle\Psi^f(\tau)|\Psi^f_{T}\rangle$ can be directly evaluated in the
GFMC propagations of $\Psi^i(\tau)$ and $\Psi^f(\tau)$ respectively.
The VMC expectation value can be cast as
\begin{eqnarray}
\langle O \rangle_V 
& = & \frac{\langle \Psi^f_T | O |\Psi^i_{T}\rangle}
           {\langle \Psi^i_T| \Psi^i_{T}\rangle} 
      \sqrt{\frac{\langle\Psi^i_{T}|\Psi^i_{T}\rangle}
           {\langle\Psi^f_{T}|\Psi^f_{T}\rangle}} \ \\ 
& = & \frac{\langle \Psi^i_T | O |\Psi^f_{T}\rangle}
           {\langle \Psi^f_T| \Psi^f_{T}\rangle} 
      \sqrt{\frac{\langle\Psi^f_{T}|\Psi^f_{T}\rangle}
           {\langle\Psi^i_{T}|\Psi^i_{T}\rangle}} \ , 
\end{eqnarray}
in which the first term can be computed in VMC walks guided by $\Psi^i_T$ and
$\Psi^f_T$, respectively.
The ratio $\langle\Psi^i_{T}|\Psi^i_{T}\rangle /
\langle \Psi^f_T| \Psi^f_{T}\rangle$ or its inverse can be also computed
in the $\Psi^i_T$ and $\Psi^f_T$ walks.
In this present GFMC calculation the propagation of the mixed off-diagonal
matrix element was generally carried out for a value of $\tau$ up to 3
MeV$^{-1}$. In most cases the transition matrix elements are quite stable with
this large value of $\tau$.

As noted in Eq.(\ref{eq:gfmc-1}), the GFMC propagation may be computed using a
simplified Hamiltonian.  In the current work we use 
$H^{\prime}$ = AV8$^{\prime}$ + IL2$^{\prime}$ where AV8$^{\prime}$
is a reprojection of AV18 defined in Ref.~\cite{PPCPW97} and
the strength of the central repulsive part of IL2 is modified in
IL2$^{\prime}$ so that 
$\langle H^{\prime} \rangle \approx \langle H \rangle$.  Energies
are then perturbatively corrected by adding 
$\langle H-H^{\prime} \rangle$.  
However all other expectation values are really for eigenfunctions
of $H^{\prime}$; we have no way of correcting them to
expectation values in the eigenfunctions of the desired $H$.
Because $H^{\prime}$ is a good approximation to $H$, this in general
should not be a problem.  However, Fermi matrix elements are
different from their trivial values ($2J+1$),
only because of charge-independence-breaking components in the
wave functions:
\begin{equation}
\langle || F || \rangle^2 = (2J+1) (1 - \epsilon) \ .
\label{f-epsilon}
\end{equation}
The AV8$^{\prime}$ does not contain the
strong charge symmetry breaking (CSB) component of AV18 and
has only a Z-dependent isoscalar projection of the Coulomb
potential as proposed by Kamuntavi\v{c}ius, et al.~\cite{GPK};
all other electromagnetic terms in AV18 are not included.
Therefore our calculations may seriously underestimate the
correct values of $\epsilon$ for AV18+IL2.  This is suggested
by a comparison with correlated hyperspherical harmonics (CHH)
calculations, using the AV18+UIX Hamiltonian,
of the F and GT matrix elements for $^3$H decay~\cite{SS+98}.  The
CHH value for the GT matrix element is 2.258 while
our result computed using AV8$^{\prime}$+UIX$^{\prime}$ is
2.260$\pm$0.001.  However their value for $\epsilon$ is 0.0013,
while ours is 0.$\pm$0.0005.

\section {Results for $A=6,7$ nuclei}
\label{numres}

Evaluation of the transition matrix elements is fairly straightforward.
The VMC or GFMC wave function samples  for a given $(J^\pi;T)$ state are
constructed with a specific $M_J$ projection.
We find it convenient to use the same $M_J$ for both initial and final
states, even if $J$ is different.
This makes the $M1$ operator exactly equivalent to the magnetic moment
operator, and also keeps the GT operator particularly simple.
We note that the size of the Monte Carlo statistical errors can vary
significantly depending on the particular $M_J$ substate that is chosen.
Below we present more than a dozen electroweak transitions between different
states of $A=6,7$ nuclei.
The first two subsections discuss the electromagnetic transitions and the
last subsection discusses the weak transitions.

\subsection{Electromagnetic Transitions of $A=6$ Nuclei}

Table~\ref{li6result} shows the VMC, the two mixed estimates, and the
extrapolated GFMC reduced matrix elements given in Eq.(\ref{eq:extrap}), for various electromagnetic
transitions between different states of $A=6$ nuclei.
The corresponding transition widths (computed with the experimental
excitation energies) are shown in Table~\ref{trwi_6}, where
they are compared with Cohen-Kurath (CK) shell-model values~\cite{CK65},
no-core shell model values (NCSM) \cite{NCSM}, and experiment~\cite{exp567,Kolata}.
The NCSM values were computed for the AV8$^\prime$+TM$^\prime$ Hamiltonian 
which we expect to have similar transitions to the AV8$^\prime$+IL2$^\prime$
Hamiltonian used here.

\begin{table}[bt]
\caption{Electromagnetic transition reduced matrix elements for $A=6$ nuclei.}
\label{li6result}
\vspace{5mm}
\begin{ruledtabular}
\begin{tabular}{ccccccc}
$J^\pi_i;T_i\to J^\pi_f;T_f$       & mode              & VMC& $M_i$& $M_f$& GFMC\\
\colrule
$^6$Li$(3^+;0)\to^6$Li$(1^+;0)$    & $E2(e~{\rm fm}^2)$ &8.20(1)~ &~~8.46(3)~~  &~~8.77(3)~  &~~9.03(5)~\\
$^6$Li$(2^+;0)\to^6$Li$(1^+;0)$    & $E2(e~{\rm fm}^2)$ &6.20(8)~ &~~6.49(16)~  &~~6.29(3)~  &~~6.58(18)\\
$^6$Li$(0^+;1)\to^6$Li$ (1^+;0)$   & $M1 (\mu_N)$      &3.682(4) &~~3.658(1)~  &~~3.643(1)  &~~3.619(4)\\
$^6$Li$(2^+;1)\to^6$Li$ (1^+;0)$   & $M1 (\mu_N)$      &0.09(3)~ &$-0.010(3)$  &$-0.010(7)$ &$-0.11(3)~$\\ 
\colrule
$^6$He$(2^+;1)\to^6$He$(0^+;1)$    & $E2(e~{\rm fm}^2)$ &1.44(1)~ &~~1.64(2)~   &~~1.63(1)~  &~~1.81(3)~
\end{tabular}
\end{ruledtabular}
\end{table}

\begin{figure}[bt]
\centerline{\epsfig{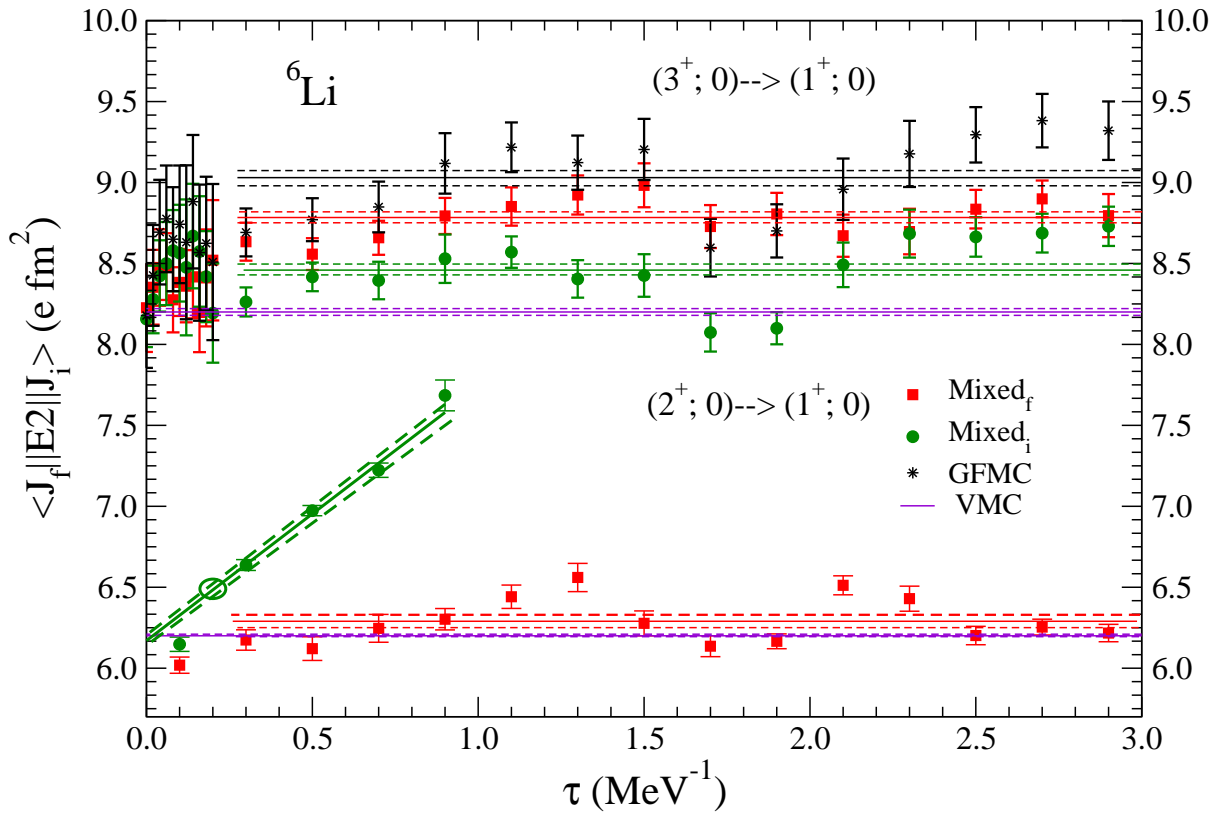}}
\caption{(Color online)
$E2$ transitions for $^6$Li$(3^+;0)$ and $^6$Li$(2^+;0)$
to $^6$Li$(1^+;0)$ ground state. \label{li6E2}}
\end{figure}

The GFMC propagation for the $^6$Li $E2$ matrix elements is shown in Fig.~\ref{li6E2}.
For each of the $E2$ transitions we plot the two mixed reduced matrix elements:
red squares for those with GFMC ground state configurations and green circles
for those with GFMC excited state configurations.
The solid purple line starting from the origin shows the pure VMC estimate, 
while the black stars
represent the extrapolated matrix elements in each $E2$ transition.
The other solid lines are the average, over the range of $\tau$ shown, 
for each reduced matrix element,
with standard deviations shown as dashed lines.

\begin{table}[bt]
\caption{Electromagnetic transition widths in eV of $A=6$ nuclei.}
\label{trwi_6}
\vspace{5mm}
\begin{ruledtabular}
\begin{tabular}{ccccccc}
$J^\pi_i;T_i\to J^\pi_f;T_f$      &      EM mode     &  CK    & NCSM\footnote{for AV8$^\prime$+TM$^\prime$ from Ref.~\cite{NCSM}} 
                                                                       &    VMC     &   GFMC      & Expt.   \\
\colrule                                                        
$^6$Li$(3^+;0)\to^6$Li$(1^+;0)$   & $E2$ ($10^{-4}$) & $2.17$ & $1.22$ & $3.86(1)~$ & $4.68(5)~$  & $4.40(34)$   \\
$^6$Li$(2^+;0)\to^6$Li$(1^+;0)$   & $E2$ ($10^{-2}$) & $0.34$ & $0.38$ & $0.92(2)~$ & $1.04(6)~$  & $0.54(28)$   \\
$^6$Li$(0^+;1)\to^6$Li$ (1^+;0)$  & $M1$ ($10^{0}$)~ & $7.85$ & $8.05$ & $7.10(2)~$ & $6.86(2)~$  & $8.19(17)$   \\
$^6$Li$(2^+;1)\to^6$Li$ (1^+;0)$  & $M1$ ($10^{0}$)~ & $0.23$ & $0.20$ & $0.003(2)$ & $0.004(2)$  & $0.27(5)~$   \\
\colrule                                                                       
$^6$He$(2^+;1)\to^6$He$(0^+;1)$   & $E2$ ($10^{-5}$) &  ---   &  ---   & $0.63(1)~$ & $1.03(3)~$  & $1.63(10)$   \\
\end{tabular}
\end{ruledtabular}
\end{table}

In Table~\ref{li6result} and Fig.\ref{li6E2} we note that for the $E2$
transition between the ground $(1^+;0)$ and the first excited $(3^+;0)$ states
of $^6$Li, the average values for the two mixed estimates are larger
than the pure VMC estimate.
As a result the extrapolated GFMC matrix element is $10\%$ larger than the
VMC value and hence the GFMC transition width is about $20\%$ larger than
the VMC width, as shown in Table~\ref{trwi_6}.
We also note that the GFMC width for this transition is a little bigger
than the experimental value, but it is within the experimental range.
It is worth mentioning that the CK shell-model prediction for this width
is about half the experimental value, despite the use of effective charges
for proton and neutron of $1.4e$ and $0.4e$, respectively.
The NCSM value is only one quarter our result, despite our expectation that the
two Hamiltonians (AV8$^\prime$+TM$^\prime$ in NCSM and AV8$^\prime$+IL2$^\prime$
in our work)
should not result in very different transition moments.  In a more recent
publication, NCSM values using a 10$\hbar\omega$ space for AV8$^\prime$ 
with no $3N$ potential were presented~\cite{NCSM2}.  The $B(E2)$ increased by 20\%
from the 6$\hbar\omega$ values presented in Table~\ref{li6result}.

\begin{figure}[bt]
\centerline{\epsfig{file=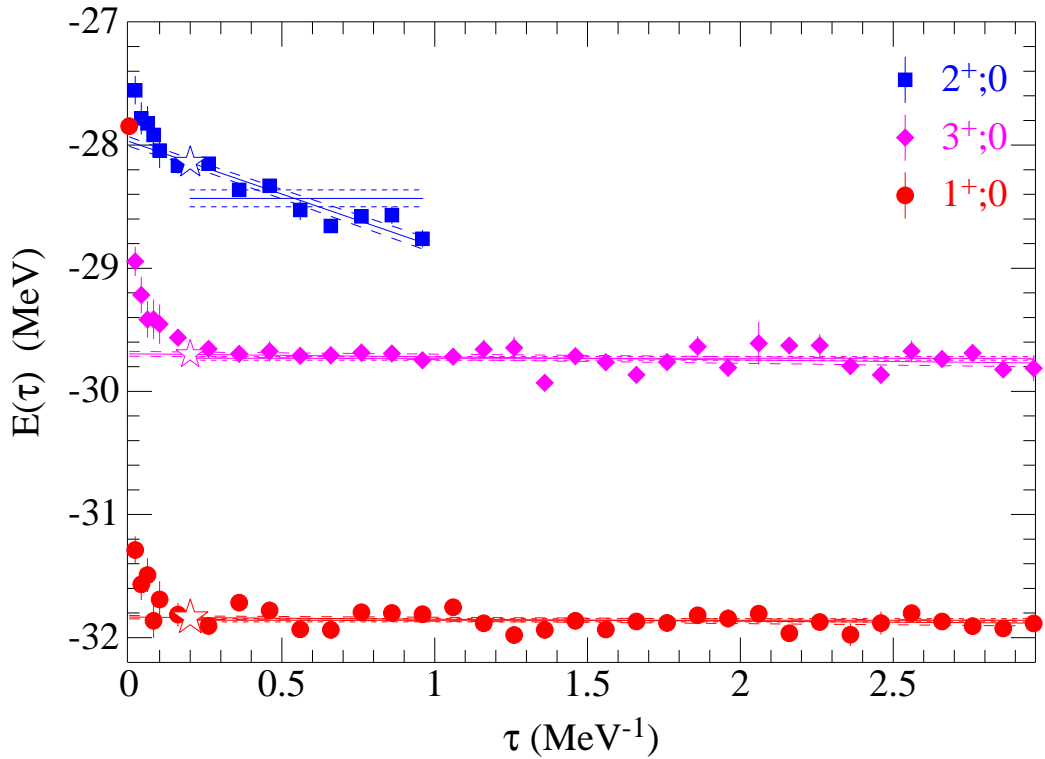,angle=-90,width=14cm}}
\caption{(Color online)
GFMC E($\tau$) for different states of $^6$Li. \label{li6e}}
\end{figure}

The $E2$ transition from $(2^+;0)$ excited state to the ground state turns out
to be difficult to calculate, as can be seen in Fig.~\ref{li6E2}.
The plot shows that the GFMC mixed estimates using either the ground or the
first excited $(3^+;0)$ states of $^6$Li nucleus are moderately stable.
On the other hand the GFMC mixed estimate which uses the broad $(2^+;0)$ state
of $^6$Li is growing rapidly with $\tau$, which makes a simple average
meaningless.
This is undoubtedly related to the difficulty in obtaining GFMC energies
for broad, particle-unstable, states.
Figure~\ref{li6e} shows the computed energies of the $(1^+;0)$, $(3^+;0)$,
and $(2^+;0)$ states in $^6$Li as a function of $\tau$.
The $(1^+;0)$ and $(3^+;0)$ energies drop rapidly with $\tau$ from the initial VMC
value and then become constant, aside from statistical fluctuations.
The stable energy is reached around $\tau=0.2$~MeV$^{-1}$ as marked
in the figure by the open stars.
However, after a similar initial rapid decrease, the energy of the
experimentally broad $(2^+;0)$ state continues to decrease.
At the same time the rms radius is steadily increasing; the GFMC algorithm
is propagating this state to separated $\alpha$ and deuteron clusters.
Based on the convergence of the $(1^+;0)$ and $(3^+;0)$ energies, we assume that
values at $\tau=0.2$~MeV$^{-1}$ represent the best GFMC estimates for this
state.
Using this value of $\tau$ for the transition results in the value 6.49(16)
for the $\langle 2^+ || E2 || 1^+ \rangle$ $M_i$ matrix element,
which is shown as an open circle in Figure~\ref{li6E2}.
The quoted error is based on a range of $\pm0.1$ for the $\tau$ at which
the value is evaluated.
This GFMC result for the width is only 10\% larger than the VMC value,
but is three times as big as the CK value, and twice as big
as the experimental value, which however has a sizeable error bar.

A similar analysis was used for the $^6$He $E2$ matrix element given in
Table~\ref{li6result}.
The experimental value here is taken from a recent measurement of the
B($E2\Uparrow$) from $^6$He breakup on $^{209}$Bi near the Coulomb
barrier~\cite{Kolata}.

\begin{figure}[bt]
\centerline{\epsfig{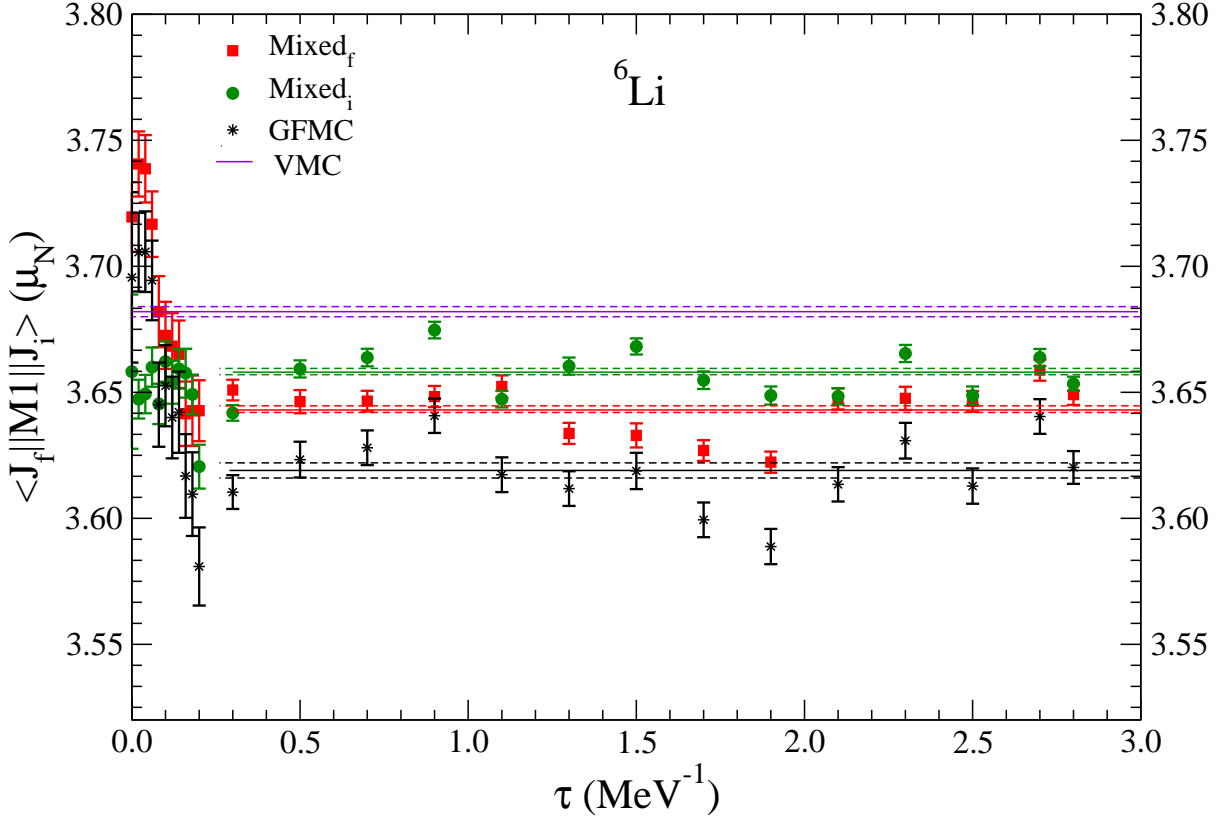}}
\caption{(Color online)
$M1$ transition for $^6$Li$(0^+;1)$ to $^6$Li$(1^+;0)$. \label{li6m1}}
\end{figure}

The GFMC propagation for the $M1$ transition $^6$Li$(0^+;1)\to^6$Li$(1^+;0)$
is shown in Fig.~\ref{li6m1}.
The GFMC matrix element reduces the VMC estimate slightly, giving a
width that is smaller than the current experimental value.
However, this is not unexpected as we have used only one-body transition
operators in our present calculation. Two-body meson-exchange currents
(MEC) are known to increase isovector magnetic moments by 15-20$\%$ for
$A=3$ nuclei, while having profound effects on the magnetic form
factors~\cite{CS98}.
A previous VMC calculation of the width of this transition found a 20\%
increase from 7.49(2) eV to 9.06(7) eV when MEC appropriate for the
AV18+UIX Hamiltonian were added~\cite{WS98}.
A similar increase applied to the present GFMC calculation would predict a
width of 8.29(3) eV in excellent agreement with the experimental width of
8.19(17) eV.
We plan to evaluate MEC corrections with the GFMC wave functions in future
work.
The NCSM result is in good agreement with experiment without any MEC
contributions; if the MEC contributions are the size we expect, however,
this good agreement will be lost when they are added.

Finally, we have a very difficult time evaluating the $M1$ transition between
$^6$Li$(2^+;1)$ and $^6$Li$(1^+;0)$ states.
The former is again a wide state which has the same GFMC propagation
difficulty as the $E2$ transition between $^6$Li$(2^+;0)$ and
$^6$Li$(1^+;0)$ states discussed above.
However, the biggest problem is a large cancellation between different
$LS[n]$ components of the two wave functions, because the dominant pieces,
$^1$D$_2[42]$ and $^3$S$_1[42]$, are not connected by the $M1$ operator.
The VMC diagonalization and GFMC propagation are driven to minimize the
energy, and may not determine small components of the wave functions
sufficiently well to obtain such sensitive cancellations.
Finally, the contribution of MEC terms may be much more important here because
of the cancellations in the impulse approximation; hints of
this were observed in the earlier VMC study~\cite{WS98}.

\subsection{Electromagnetic Transitions of $A=7$ Nuclei}

In Table~\ref{trme_7} we present the matrix elements of a number of
electromagnetic transitions in $A=7$ nuclei.
As in Table~\ref{li6result}, this table shows the VMC estimates, the two
mixed estimates and the GFMC extrapolated matrix elements for each transition.
We suppress the isospin quantum numbers for different states of $^7$Li and 
$^7$Be because all states we consider have $T=\frac{1}{2}$. 
For those transitions in $A=7$ nuclei between particle-stable states, we
made two independent calculations using both Type~I and Type~II trial wave
functions as discussed in Sec.~\ref{sec:VMCwaves}.
However, the Type~II trial function is not defined for particle-unstable 
states like $^7$Li($\frac{7}{2}^-$).
It is expected that, even though the VMC estimates may be somewhat dependent
on the trial wave functions, the GFMC calculation should remove most of the
dependence.  (It is exact at the order of $|\Psi_0-\Psi_T|^2$).
The extrapolated regular (I) and asymptotic (II) expectation values are within
$2\%$ of each other for every transition we considered.

\begin{table}[bt]
\caption{Electromagnetic transition reduced matrix elements of $A=7$
nuclei.\label{trme_7}}
\vspace{5mm}
\begin{ruledtabular}
\begin{tabular}{crcccccc}
$J^\pi_i\to J^\pi_f$                        & $\Psi_T$ &              mode & VMC      & $M_i$   & $M_f$   & GFMC\\
\colrule
$^7$Li$(\frac{1}{2}^-) \to ^7$Li$(\frac{3}{2}^-)$ &  I & $E2(e~{\rm fm}^2)$ &5.11(5)~  &5.44(2)~ &5.37(2)~ &~5.69(6)~ \\
$^7$Li$(\frac{1}{2}^-) \to ^7$Li$(\frac{3}{2}^-)$ & II & $E2(e~{\rm fm}^2)$ &5.38(6)~  &5.53(3)~ &5.56(2)~ &~5.71(7)~ \\
$^7$Li$(\frac{1}{2}^-) \to ^7$Li$(\frac{3}{2}^-)$ &  I & $M1(\mu_N)$       &2.742(1)  &2.749(3) &2.693(2) &~2.695(4) \\
$^7$Li$(\frac{1}{2}^-) \to ^7$Li$(\frac{3}{2}^-)$ & II & $M1(\mu_N)$       &2.738(1)  &2.673(3) &2.706(2) &~2.641(3) \\
$^7$Li$(\frac{7}{2}^-) \to ^7$Li$(\frac{3}{2}^-)$ &  I & $E2(e~{\rm fm}^2)$ &7.67(4)~  &8.28(3)~ &8.30(3)~ &~8.91(6)~ \\
\colrule                                           
$^7$Be$(\frac{1}{2}^-) \to ^7$Be$(\frac{3}{2}^-)$ &  I & $E2(e~{\rm fm}^2)$ &8.51(3)~  &9.09(4)~ &9.54(3)~ &10.12(6)~ \\
$^7$Be$(\frac{1}{2}^-) \to ^7$Be$(\frac{3}{2}^-)$ & II & $E2(e~{\rm fm}^2)$ &8.86(13)  &9.74(7)~ &9.60(6)~ &10.48(16) \\
$^7$Be$(\frac{1}{2}^-) \to ^7$Be$(\frac{3}{2}^-)$ &  I & $M1(\mu_N)$       &2.423(2)  &2.412(2) &2.403(3) &~2.394(4) \\
$^7$Be$(\frac{1}{2}^-) \to ^7$Be$(\frac{3}{2}^-)$ & II & $M1(\mu_N)$       &2.405(3)  &2.390(2) &2.386(5) &~2.372(6) \\
\end{tabular}
\end{ruledtabular}
\end{table}

Table~\ref{trwi_7} shows the corresponding widths of the electromagnetic transitions
in $A=7$ nuclei, compared to the CK shell-model values~\cite{CK65} and
experiment~\cite{exp567,li7}.
The GFMC width is about 20\% bigger than the CK and VMC values, and in
good agreement with the experimental width for the
$^7$Li$(\frac{1}{2}^-) \to ^7$Li$(\frac{3}{2}^-)$ $E2$ transition.
The corresponding transition in $^7$Be has not been measured.
The widths for both the $M1$ transitions are relatively smaller than the
corresponding experimental widths as expected for just one-body
magnetic-moment operator expectation values.
If there is a 20\% additional contribution as expected from MEC
terms, these will approach within 10\% of the experimental values.

\begin{table}[bt]
\caption{Electromagnetic transition widths in eV of $A=7$ nuclei.
\label{trwi_7}}
\vspace{5mm}
\begin{ruledtabular}
\begin{tabular}{crccccc}
$J^\pi_i\to J^\pi_f$                      & $\Psi_T$ & EM mode          &  CK  &    VMC   &   GFMC   & Expt.   \\
\colrule
$^7$Li$(\frac{1}{2}^-)\to^7$Li$(\frac{3}{2}^-)$ &  I & $E2$ ($10^{-7}$) & 2.79 & 2.61(3)  & 3.24(7)~ & 3.30(20)\\
$^7$Li$(\frac{1}{2}^-)\to^7$Li$(\frac{3}{2}^-)$ & II & $E2$ ($10^{-7}$) & ---  & 2.90(3)  & 3.26(8)~ & 3.30(20)\\
$^7$Li$(\frac{1}{2}^-)\to^7$Li$(\frac{3}{2}^-)$ &  I & $M1$ ($10^{-3}$) & 5.69 & 4.74(3)  & 4.58(3)~ & 6.30(31)\\
$^7$Li$(\frac{1}{2}^-)\to^7$Li$(\frac{3}{2}^-)$ & II & $M1$ ($10^{-3}$) & ---  & 4.73(1)  & 4.41(1)~ & 6.30(31)\\
$^7$Li$(\frac{7}{2}^-)\to^7$Li$(\frac{3}{2}^-)$ &  I & $E2$ ($10^{-2}$) & 0.98 & 1.29(1)  & 1.74(2)~ & 1.50(20)\\
\colrule                                                                                    
$^7$Be$(\frac{1}{2}^-)\to^7$Be$(\frac{3}{2}^-)$ &  I & $E2$ ($10^{-7}$) & ---  & 4.24(3)~ & 6.00(7)~ & ---\\
$^7$Be$(\frac{1}{2}^-)\to^7$Be$(\frac{3}{2}^-)$ & II & $E2$ ($10^{-7}$) & ---  & 4.60(13) & 6.44(19) & ---\\
$^7$Be$(\frac{1}{2}^-)\to^7$Be$(\frac{3}{2}^-)$ &  I & $M1$ ($10^{-3}$) & ---  & 2.69(1)~ & 2.62(1)~ & 3.43(45)\\
$^7$Be$(\frac{1}{2}^-)\to^7$Be$(\frac{3}{2}^-)$ & II & $M1$ ($10^{-3}$) & ---  & 2.65(1)~ & 2.57(1)~ & 3.43(45)\\
\end{tabular}
\end{ruledtabular}
\end{table}

\begin{figure}[bt]
\centerline{\epsfig{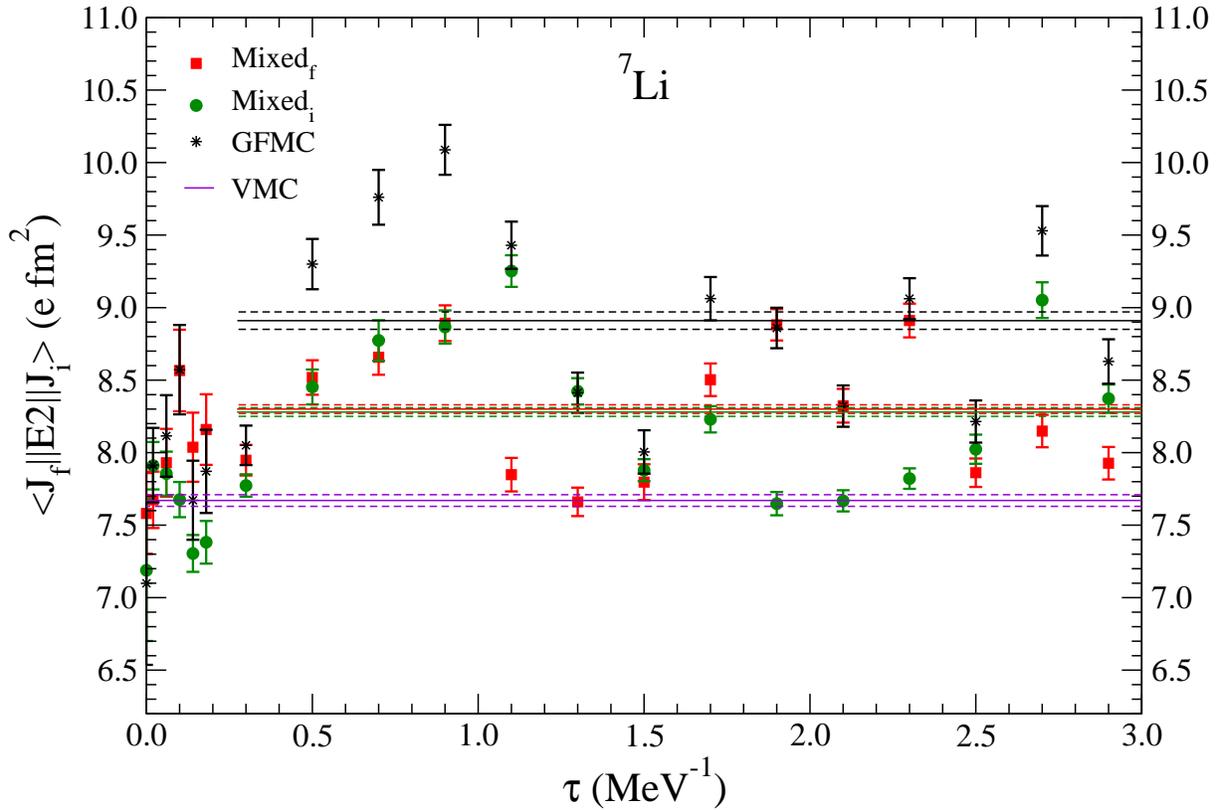}}
\caption{(Color online)
$E2$ transition for $^7$Li$(\frac{7}{2}^-)$ to $^7$Li$(\frac{3}{2}^-)$.
\label{E2li7to3}}
\end{figure}

The GFMC propagation for the $E2$ transition from $^7$Li$(\frac{7}{2}^-)$ to
$^7$Li$(\frac{3}{2}^-)$ is illustrated in Figure~\ref{E2li7to3}.
We note that even though the individual points for the two mixed estimates
are a little scattered, the average values for these overlap.
The extrapolated GFMC result is larger than the VMC estimate by 15\%,
making the transition width 30\% larger.
The VMC value is one experimental standard deviation below the experimental value~\cite{li7}
while the GFMC value is the same amount above the experimental value.

\subsection{Weak Transitions in $A=6,7$ nuclei}

The weak Fermi and Gamow-Teller matrix elements in $A=6,7$ nuclei are shown in
Table~\ref{trme_67}.
We note that in every F and GT transition the extrapolated GFMC matrix elements
are smaller than the VMC estimates.
However, the reduction is not large, being about $2\%$ for GT terms.
This suggests the starting trial functions are already good approximations
for these weak transitions.
The differences between Type~I and Type~II trial functions are not great,
and the GFMC propagation does not reduce these differences.

The Fermi matrix element for $^7$Be$(\frac{3}{2}^-)$ to $^7$Li$(\frac{3}{2}^-)$ is exactly
2 for charge-symmetric wave functions such as our Type~I.
The Type~II wave functions are not charge-symmetric because the $\alpha t$
and $\alpha \tau$ separation energies in Eq.(\ref{eqn:type2}) are different
and also the triton and $^3$He clusters are slightly different.
It appears that the GFMC propagation will introduce only a small asymmetry
when starting from a charge-symmetric (Type~I) trial function, but if given
a small starting asymmetry (Type~II), it can enhance it considerably.
However, as noted at the end of Sec. III, the present calculations
may still seriously underestimate this asymmetry and its
effect on $\langle || F || \rangle$.
We also show in the last line of Table~\ref{trme_67} the GT matrix element 
for the transition from $^7$Be($\frac{1}{2}^-$) to $^7$Li($\frac{3}{2}^-$)
which is not an observable weak decay, but could be measured in a ($p,n$)
reaction.  The charge-symmetric Type~I trial function would give the
same result as the $^7$Be($\frac{3}{2}^-$) to $^7$Li($\frac{1}{2}^-$)
transition, but the Type~II trial function gives a very slightly different 
result.

\begin{table}[bt]
\caption{Weak transition reduced matrix elements of $A=6,7$ nuclei.}
\label{trme_67}
\vspace{5mm}
\begin{ruledtabular}
\begin{tabular}{crccccc}
$J^\pi_i\to J^\pi_f$                         & $\Psi_T$ & mode& VMC       & $M_i$    & $M_f$    & GFMC \\
\colrule
$^6$He$(0^+;1) \to^6$Li$(1^+;0)$                  &  I  & GT  &2.195(1)~  &2.176(1)~ &2.175(1)~ &2.157(1)~ \\
$^6$He$(0^+;1) \to^6$Li$(1^+;0)$                  & II  & GT  &2.253(3)~  &2.184(1)~ &2.276(1)~ &2.207(3)~ \\
\colrule                                            
$^7$Be$(\frac{3}{2}^-) \to ^7$Li$(\frac{3}{2}^-)$ &  I  & F   &2.0000(0)  &1.9998(1) &1.9998(1) &1.9997(3) \\
$^7$Be$(\frac{3}{2}^-) \to ^7$Li$(\frac{3}{2}^-)$ & II  & F   &1.9995(1)  &1.9987(3) &1.9983(3) &1.9976(5) \\
$^7$Be$(\frac{3}{2}^-) \to ^7$Li$(\frac{3}{2}^-)$ &  I  & GT  &2.325(1)~  &2.298(1)~ &2.301(1)~ &2.274(2)~ \\
$^7$Be$(\frac{3}{2}^-) \to ^7$Li$(\frac{3}{2}^-)$ & II  & GT  &2.339(4)~  &2.311(1)~ &2.319(1)~ &2.291(4)~ \\
$^7$Be$(\frac{3}{2}^-) \to ^7$Li$(\frac{1}{2}^-)$ &  I  & GT  &2.146(2)~  &2.119(3)~ &2.129(3)~ &2.099(4)~ \\
$^7$Be$(\frac{3}{2}^-) \to ^7$Li$(\frac{1}{2}^-)$ & II  & GT  &2.139(1)~  &2.121(1)~ &2.098(2)~ &2.080(1)~ \\
$^7$Be$(\frac{1}{2}^-) \to ^7$Li$(\frac{3}{2}^-)$ & II  & GT  &2.138(1)~  &2.125(3)~ &2.104(1)~ &2.092(3)~ \\
\end{tabular}
\end{ruledtabular}
\end{table}

The log($ft$) values obtained from VMC, shown in Table~\ref{trwi_67}, are
already in reasonable agreement with the corresponding experimental values
and the GFMC values are even better.
The previous VMC study~\cite{SW02} included MEC contributions which boosted
the GT transition matrix elements $\sim 1.5\%$ for $A=6$, and $\sim 3\%$
for $A=7$.
This resulted in too small a half-life for $^6$He but about right for $^7$Be.
When MEC contributions are eventually added to the GFMC calculation,
the half-life for $^6$He should be quite good, but the rate for $^7$Be
will probably be a little too fast.
The last two lines of Table~\ref{trwi_67} give the branching ratio $\xi$ of the
weak decay to the two final states in $^7$Li for Type I and II trial 
functions.  These are also a little low compared to experiment, but MEC 
contributions should also improve the agreement.

\begin{table}[ht]
\caption{Log($ft$) values for weak transitions of $A=6,7$ nuclei}
\label{trwi_67}
\vspace{5mm}
\begin{ruledtabular}
\begin{tabular}{crcccccc}
 $J^\pi_i\to J^\pi_f$                    & $\Psi_T$ &Weak Current &  CK   & NCSM\footnote{for AV8$^\prime$+TM$^\prime$ from Ref.~\cite{NCSM}} 
                                                                                         &    VMC   &   GFMC   & Expt.   \\
\colrule
$^6$He$(0^+;1) \to$  $ ^6$Li$(1^+;0)$                &  I  & GT      & 2.84   & 2.87 & 2.901(1)~ & 2.916(1)~ & 2.910(2) \\
$^6$He$(0^+;1) \to$  $ ^6$Li$(1^+;0)$                & II  & GT      & ---    & ---  & 2.879(2)~ & 2.897(2)~ & ---      \\
\colrule                                                                                        
$^7$Be$(\frac{3}{2}^-)\to^7$Li$(\frac{3}{2}^-)$  &  I  & F \& GT & 3.38   & 3.30 & 3.288(1) & 3.302(1) & 3.32 \\
$^7$Be$(\frac{3}{2}^-)\to^7$Li$(\frac{3}{2}^-)$  & II  & F \& GT & ---    & ---  & 3.285(1) & 3.297(1) & ---  \\
$^7$Be$(\frac{3}{2}^-)\to^7$Li$(\frac{1}{2}^-)$  &  I  & GT      & 3.46   & 3.53 & 3.523(1) & 3.542(1) & 3.55 \\
$^7$Be$(\frac{3}{2}^-)\to^7$Li$(\frac{1}{2}^-)$  & II  & GT      & ---    & ---  & 3.526(1) & 3.550(1) & ---  \\
$\xi$ Li$(\frac{1}{2}^-)/$Li$(\frac{3}{2}^-)$    &  I  & F \& GT & 14.2\% & 10.38\% & 10.38(3)\%  & 10.25(3)\%  & 10.44(4)\% \\
$\xi$ Li$(\frac{1}{2}^-)/$Li$(\frac{3}{2}^-)$    & II  & F \& GT & ---    & ---     & 10.25(3)\%  & 10.00(3)\%  & ---        \\  
\end{tabular}
\end{ruledtabular}
\end{table}

\begin{figure}[bt]
\centerline{\epsfig{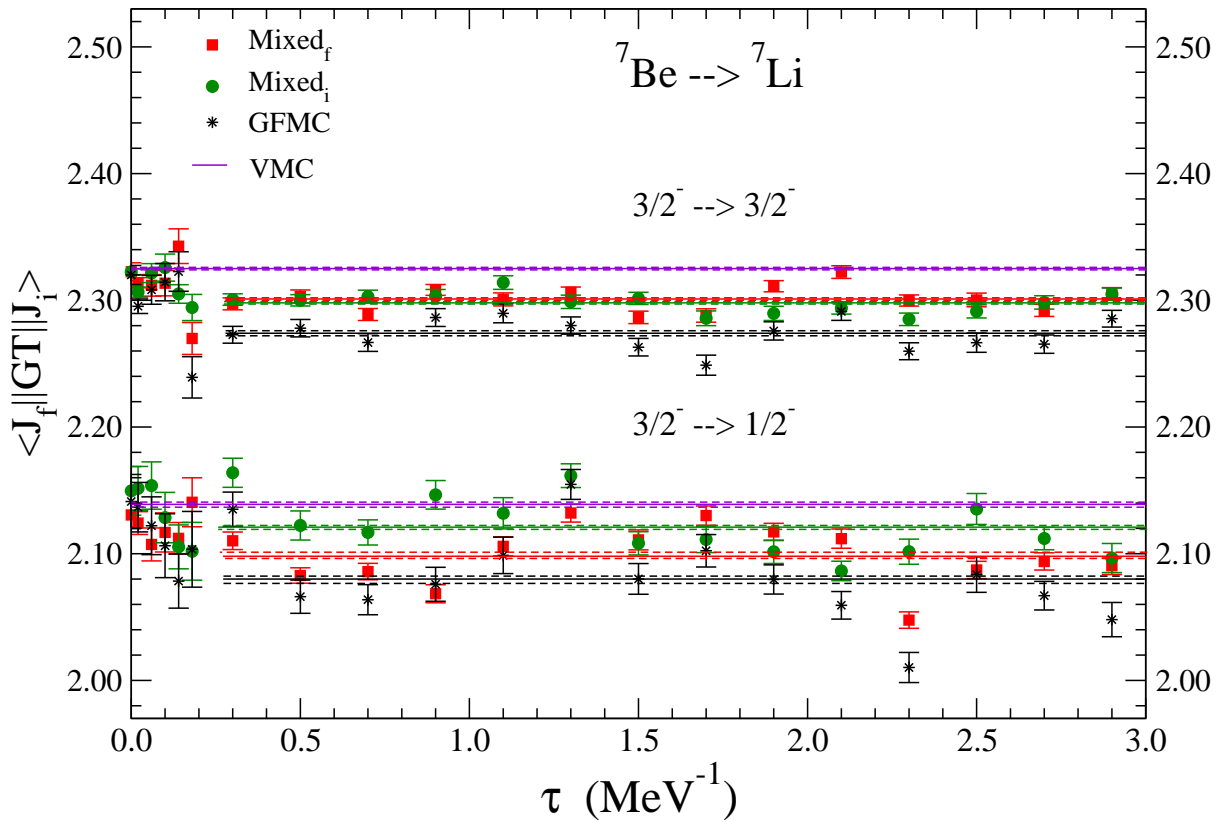}}
\caption{(Color online)
Gamow-Teller transitions for $^7$Be$(\frac{3}{2}^-)$ to $^7$Li$(\frac{3}{2}^-)$
and $^7$Be$(\frac{3}{2}^-)$ to $^7$Li$(\frac{1}{2}^-)$. \label{gtbe3toli3a1}}
\end{figure}

In Fig.~\ref{gtbe3toli3a1} we present the various reduced matrix elements as a
function of $\tau$ for two Gamow-Teller transitions,
$^7$Be$(\frac{3}{2}^-)$ to $^7$Li$(\frac{3}{2}^-)$ and $^7$Be$(\frac{3}{2}^-)$ to $^7$Li$(\frac{1}{2}^-)$.
The former is shown for the Type~I trial function, and the latter for the
Type~II.
The GFMC mixed estimate points for both transitions are quite stable.

\section {Conclusions}

These first GFMC calculations of transition matrix elements in light
$A=6,7$ nuclei are generally in good agreement with the current
experimental data.
A number of these transitions have been calculated previously using the more
approximate VMC technique with an older $3N$ potential~\cite{WS98,LW01,SW02}.
Here we explored a significant number of electroweak transitions using the
GFMC method, and calculated the corresponding widths or log($ft$) for each
transition.
We compared our results to Cohen-Kurath shell-model and no-core shell-model 
results where available, and also
with the current experimental results.
In most of the transitions we considered we found that the GFMC transition
widths or log($ft$) values have been improved from the VMC and are in good
agreement with experimental numbers.
This is in general true for most of the $E2$ and all the GT and F type
transitions.
However, for $M1$ type transitions the GFMC widths we obtained are smaller
than the current experimental values.
Meson-exchange current corrections are expected to be large for $M1$
transitions and must be calculated for a meaningful comparison with data.
We note here that the effect of MEC on $E2$ and GT transitions are expected
to be smaller, about $\leq 3\%$.
In addition to the good results we obtained in most cases we faced some
difficulties, especially treating broad nuclear states using the GFMC method.
In these cases scattering boundary conditions should be used; GFMC has
recently been successfully applied to the $n \alpha$ scattering
states~\cite{NPWCH07}.
We also had difficulty when the main components of the wave functions did
not contribute to the transition, with the result depending on cancellations
between small components.

Some of the transitions that we explored were also treated by using two
different types of trial wave functions.
The extrapolated GFMC values
obtained by using either of the wave functions should be the same;
in practice they are within 2\% of each other.
This is indeed found in our calculations; the widths we obtain by using
one or the other type of trial wave function are very close.

In future, we expect to extend this work to larger nuclei in the
$A$=8-10 range and for additional operators such as $E1$ and $M3$.
One difficulty we anticipate is that some transitions of interest, such
as the weak decays of $^8$He, $^8$Li, and $^8$B, run predominantly from
large components of the initial state wave function to small components
in the final states.
These small components may not be well-determined by the GFMC calculation
so additional constraints may be necessary.
We also need to evaluate two-body contributions to the electroweak
current operators that are consistent with our chosen Hamiltonian.

\acknowledgments

The many-body calculations were performed on the parallel computers of the
Laboratory Computing Resource Center, Argonne National Laboratory.
This work is supported by the U. S. Department of Energy,
Office of Nuclear Physics, under contract No. DE-AC02-06CH11357
and under SciDAC grant No. DE-FC02-07ER41457.

\end{document}